\documentclass[aps,epsfig,preprint]{revtex4-1}
\usepackage{color} 
\usepackage{xcolor}
\usepackage{graphicx} 
\usepackage{epsfig}
  \usepackage{kpfonts}

\newcommand{\bee}{\begin{equation}}
\newcommand{\ene}{\end{equation}}
\newcommand{\beea}{\begin{eqnarray}}
\newcommand{\enea}{\end{eqnarray}}

\begin{document}
\title{Merger and reconnection  of Weibel separated relativistic electron beam}
 \author{Chandrasekhar Shukla}
 \email{chandrasekhar.shukla@gmail.com}
 \author{Atul Kumar}
 \author{Amita Das}
 \email{amita@ipr.res.in}
 \author{Bhavesh Patel}
 \affiliation{Institute for Plasma Research, HBNI, Bhat, Gandhinagar - 382428, India }
 
\date{\today}
\begin{abstract} 
The relativistic electron beam (REB)  propagation in a plasma is fraught with beam plasma instabilities.  The prominent amongst them being the 
collisionless Weibel destabilization which spatially separates the   forward propagating REB 
 and the  return shielding currents. This results in the formation of REB current filaments which are typically of  the size of electron 
skin depth during the linear stage of the instability. It has been observed  that in the nonlinear stage the filaments size increases as they merge with each other. 
With the help of 2-D PIC simulations in the plane perpendicular to the REB propagation,  it is shown  that these mergers occur in   two distinct nonlinear phases.  In the first phase, the total magnetic energy increases. 
 Subsequently, however, during  the second phase, one observes  a reduction  in magnetic energy. 
It is shown that the  transition from one nonlinear regime  to another occurs when the typical 
 current associated with  individual filaments hits the Alfv\'en threshold. In the second   nonlinear regime, therefore,  the filaments can no longer 
 permit any increase in current.  Magnetic reconnection events then dissipate the excess current (and its associated magnetic energy)
  that would result from a merger process   leading to the generation of energetic electrons jets in the perpendicular plane. 
  At later times when there are only few filaments left 
 the individual reconnection events can be clearly identified. It is observed that in between such events the magnetic energy remains constant and 
 shows a sudden drop as and when two filaments merge. The electron jets released in these reconnection events are thus responsible for the transverse heating 
which has been  mentioned in some previous  studies [Honda et al. Phys. Plasmas 7, 1302 (2000)].  

\end{abstract}
 \maketitle 
\section{INTRODUCTION}
The existence and impact of the magnetic field in astrophysical events have continued to excite researchers, positing interesting issues pertaining to plasma physics. 
With the advent of high intensity lasers, it has been possible to make  interesting observations on the dynamical evolution of magnetic field  in laboratory experiments 
on laser matter interaction  
\cite{Stamper,Fujioka,mondal,Flacco,gaurav}. The intense lasers ionize the matter into plasma state and dump their energy into the lighter 
electron species, generating relativistic electron beam (REB) \cite{Modena,malka,joshi}  in the medium.  Though the 
 propagation of relativistic electron beam with current more than Alfv\'en current limit i.e. $I= (m_ec^{3}/e)\gamma_b = 17{\beta\gamma_
{b}} kA$, where $\beta=v_b/c$, $v_b$ is velocity of beam  and $\gamma_b=(1-v_b^2/c^2)^{-1}$ is relativistic Lorentz factor, is not permitted in the vacuum ( as the associated magnetic fields are large enough to totally curve back the trajectories of the electrons).
In plasma medium, this is achieved as the 
 current due to REBs are compensated by the return shielding current in the opposite direction 
provided by the electrons of the background plasma medium. 
The two currents initially  overlap spatially, resulting in zero net currents and so  no magnetic field is present initially. 
The combination of forward and reverse shielding current is, however, susceptible to   several micro-instabilities. A leading instability in the  
relativistic regime is the filamentation instability \cite{fried}.  It is often also termed as the Weibel instability \cite{weibel}. 
 The  filamentation/ Weibel instability creates  spatial separation of the forward and reverse shielding currents.  
 The current separation leads to the generation of the magnetic 
field at the expense of the kinetic energy of the beam and plasma particles. 
 The typical scale length at which the Weibel separation has the maximum growth rate is at  the electron skin depth scale 
$c/\omega_{pe}$. The Weibel separation, thus,  leads to the formation of REB current filaments of the size of  electron skin 
depth scale and is responsible for the growth of the magnetic  energy in the system. 
The dynamics, long term evolution and energetics associated with the  Weibel instability of current filaments 
are of central importance in many contexts. 
 For instance, in  fast ignition concept of fusion, the energetic REB is expected to create an ignition spark at the 
 compressed core of the target for which it has to  traverse the lower density plasma corona \cite{tabak_05,john,taguchi,hill,honda} and dump its energy 
 at the central dense core of the target. This requires a complete understanding of REB propagation in the plasma medium. 
In astrophysical scenario, the generation of cosmological magnetic field and relativistic collisionless shock formation
in gamma ray bursts  have often been attributed to the  collisionless Weibel instability \cite{cosmaggen,Huntington,med1,med2,silva}. 
The formation of collisionless shock and it's behavior depends 
on long term evolution and dynamics of the  magnetic field generated through Weibel destabilization process. 

\begin{figure}
       \centering
               \includegraphics[width=0.65\textwidth]{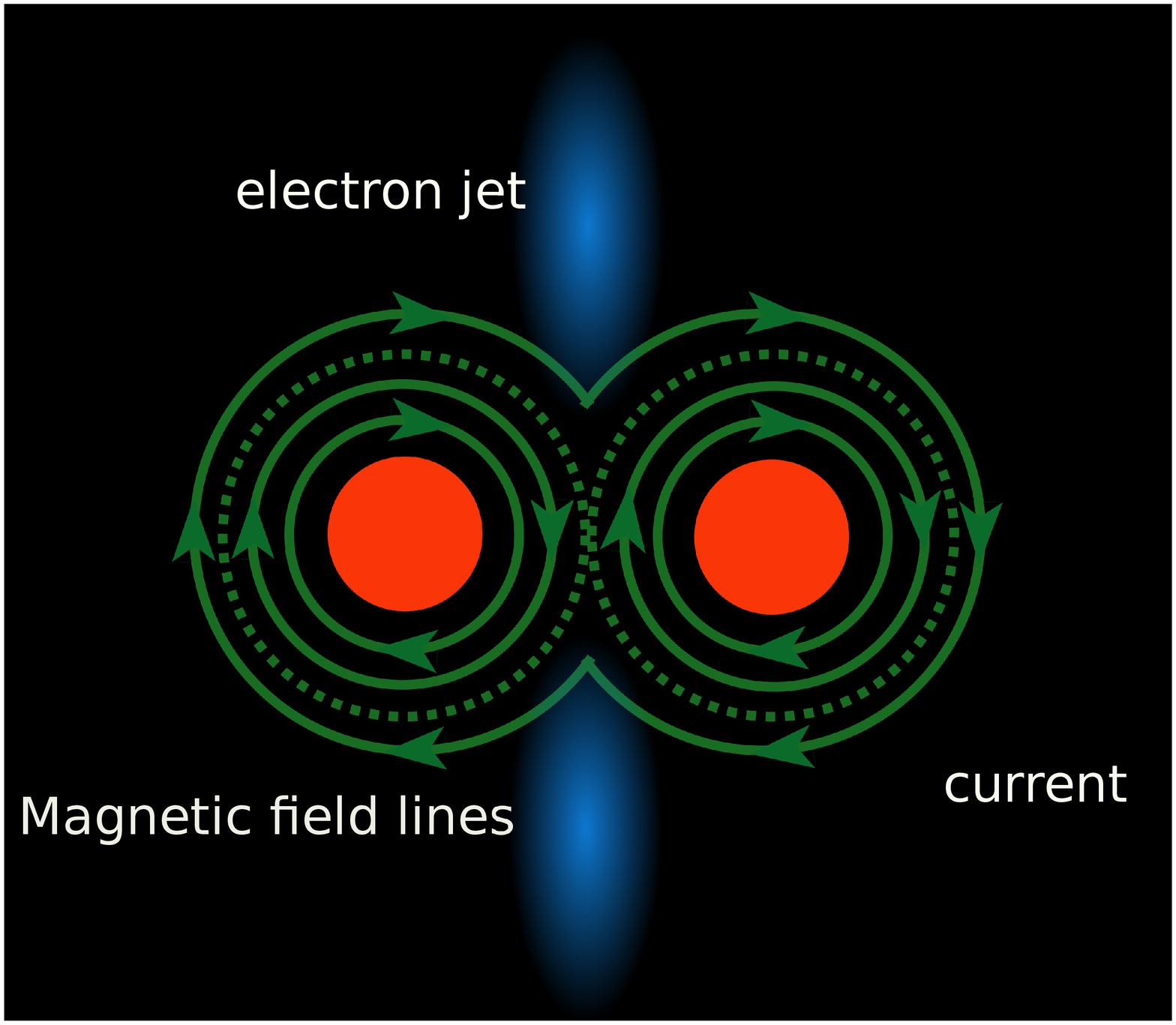} 
                \caption {Schematic of magnetic reconnection where the magnetic field lines reconnects and accelerates the plasma particles as a jet  }
                \label{fig:schematic}
       \end{figure}

The growth of  magnetic field  through the Weibel destabilization process influences the 
 propagation of REB filaments.  In this nonlinear stage, the current filaments are observed to coalesce  
 and  form the larger structure.  There are indications from previous studies \cite{polo} that at the early nonlinear stage 
 of evolution there is a growth of magnetic field energy. Subsequently, however, the  magnetic energy shows decay. 
 The physical mechanism for the observed  decay of magnetic field at later stages is the focus of the present studies. There are suggestions that 
  the merging process of super Alfv\'enic currents carrying filaments leads to the decay of magnetic energy \cite{polo}. On the other hand, the mechanism of  magnetic reconnection  
  which rearranges the magnetic topology in plasma 
  is also invoked which converts the  magnetic energy
 to kinetic energy of the particles. This can result in  thermal particles  or  the particles   may even get  accelerated \cite{recon1,recon2} by the reconnection process. 
 The merger of current filaments leading to $X$ point formation where reconnection happens 
 has been shown in the schematic diagram  Fig.~\ref{fig:schematic}.

 In this work, we have studied the linear  and non-linear stage of Weibel instability in detail with the help of   2-D Particle - In- Cell (PIC) simulation.   
 The 2-D plane of simulation is perpendicular to the current flow direction.  The initial condition is chosen as two overlapping 
 oppositely propagating electron currents. The development of the instability, 
 the characteristic features during  nonlinear phase etc., are studied in detail.

 The paper has been organized as follows. The simulation set up   has been  discussed in section II. The observations corresponding to the 
 linear phase of instability is presented in section III. 
 The nonlinear phase of the instability covered  in section IV. Section V provides for the summary and the discussion.

\section{SIMULATION SET-UP}
We employ OSIRIS2.0 \cite{os1, os2}  Particle - In - Cell (PIC) code to study the evolution of the two counterstreaming 
electron current flows in a 2-D $x1-x2$ plane perpendicular to the current flow direction of $\pm \hat{z}$. 
We have considered the ion response to be negligible and treated them as merely providing a stationary neutralizing background. 
Thus the dynamics is governed by electron species alone. However, this would not be applicable at longer times where ion response 
may become important and introduce new features.

The boundary conditions are chosen to be  periodic for both the
electromagnetic field and the charged particles in all direction. We choose the  area of the simulation box $R$  as $64\times64$ $(c/\omega_{pe})^2$ corresponding to 
$640\times640$ cells. The time   step is chosen  to be $7.07\times 10^{-2}/\omega_{pe}$ where $\omega_{pe}=\sqrt{4\pi n_{0e}e^2/m_e}$ and $n_{0e}=n_{0b}+n_{0p}$ 
is the total electron density which is the sum of beam and the  plasma electrons denoted by suffix $b$ and $p$ respectively. 
The total number of electrons and ions per cell in the  simulations are chosen to be $500$ each.
The quasi neutrality  is maintained  in the system by 
choosing equal number of electrons and ions. 
The    velocity of beam electrons and the cold plasma electrons are chosen to satisfy  current neutrality condition.
The uniform plasma density $n_{0e}$ is taken as $1.1\times 10^{22} cm^{-3}$ and 
the ratio of electron beam density to background electron density has been taken  as($n_{0b}/n_{0p}=1/9$) is the simulations presented here. 
The fields are normalized by $m_ec\omega_{pe}/e$. The evolution of field energy normalized by $m_{e}c^2n_{0e}$ is averaged over the simulation box. 
We have carried simulations for the choice of cold as well as finite temperature beams. Several choices of beam temperature were considered. 

\section{LINEAR STAGE OF INSTABILITY}
\begin{figure}
       \centering
               \includegraphics[width=0.9\textwidth]{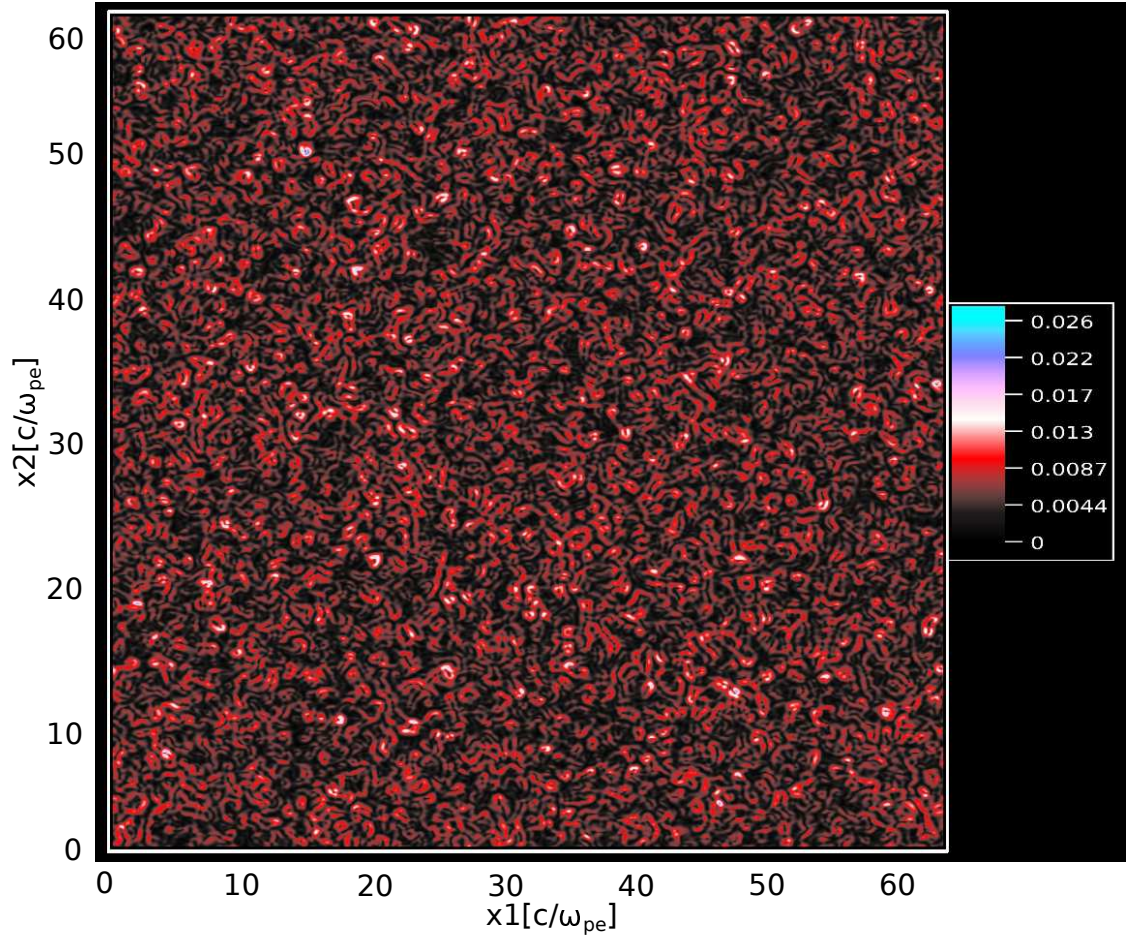} 
                \caption {The transverse magnetic field $B_{\perp}=\sqrt{B_{x}^2+B_{y}^2}$[in unit of $m_ec\omega_{pe}/e$] at $t\omega_{pe}=51$: The size of magnetic field structure in linear regime is the order of $c/\omega_{pe}$}
                 \label{fig:mag_theta}
       \end{figure}

The charge neutrality and the current balance condition, chosen initially ensures that there are no electric and magnetic fields associated with the system 
and equilibrium conditions are satisfied. We observe a development of magnetic field structures of the typical size of electron skin depth 
with time. This can be seen in Fig.~\ref{fig:mag_theta} where the contours of the transverse magnetic field have been shown in the 2D $x1-x2$ plane. The development of this magnetic field 
can be understood as arising  due to the 
spatial separation of forward and return currents through Weibel destabilization process. The growth of transverse magnetic field energy with time has been shown in Fig.3 for the 
two cases with following parameters: (I) $v_{0b}=0.9c$ and (II) $v_{0b}=0.9c$, $T_{0b}=1kev$, $T_{0p}=0.1kev$.  
The initial linear phase of growth is depicted by the straight line region in the log linear plot. A comparison with the analytically estimated  maximum growth rate for the two cases 
 has been provided by the dashed line drawn alongside. 

\begin{figure}
               \includegraphics[width=\textwidth]{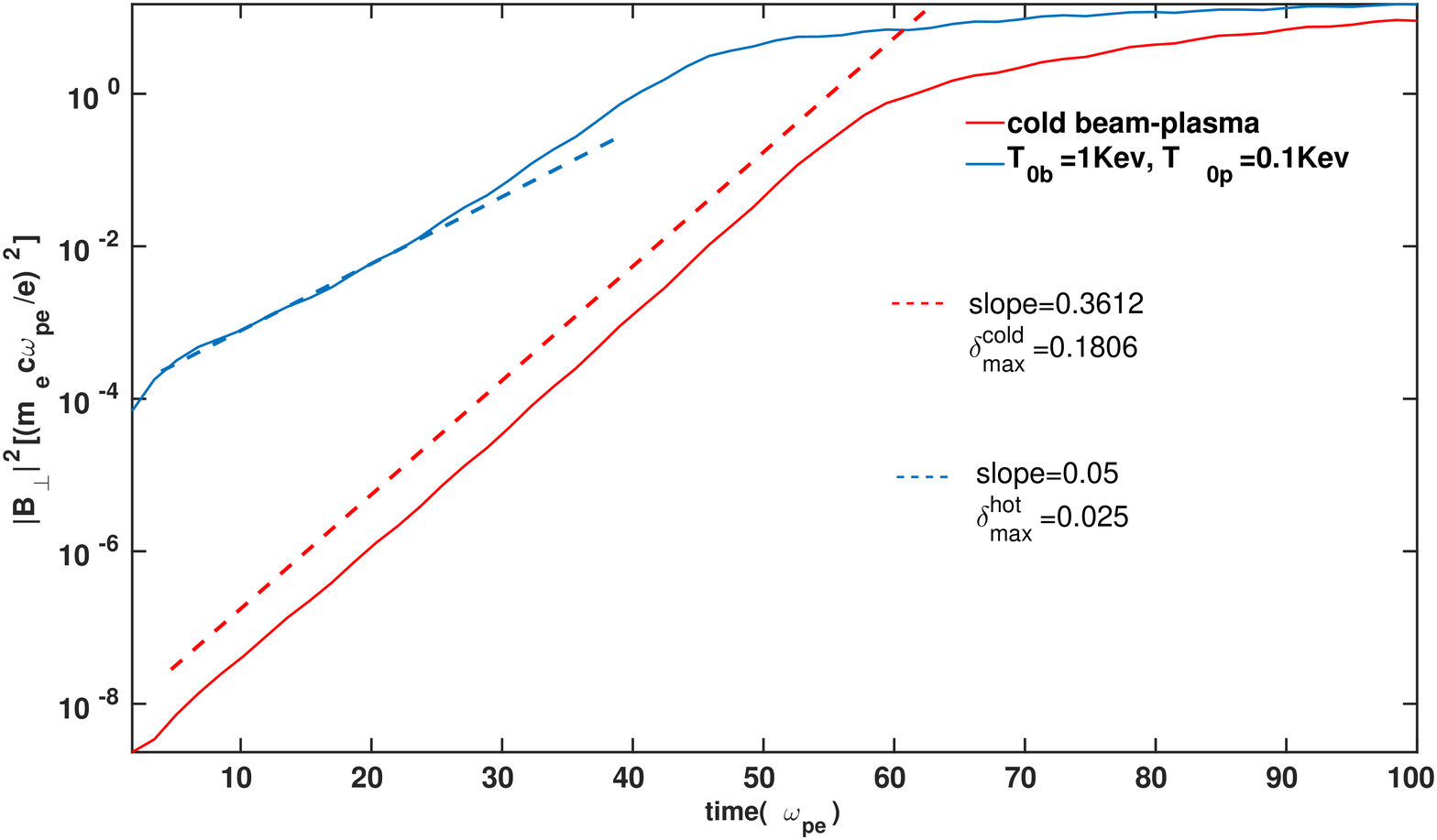} 
                \caption{Calculation of linear growth rate from PIC simulation  }  
                \label{fig:l_growth_Rate}
        \end{figure}

 The growth rate for case  (I) with $v_{0b}=0.9c$ the growth rate obtained from the simulation by measuring the half of the slope of the magnetic energy evolution in Fig.~\ref{fig:l_growth_Rate} is $0.18$ 
 and it compares well with the analytical value of 
 $0.1879$ ($ \delta^{cold}_{max}\sim \left(\frac{v_{0b}}{c}\right)\sqrt{\frac{n_{0b}/n_{0p}}{\gamma_{0b}}}\omega_{p}$\cite{Godfrey}). Similarly for case (II) when the beam temperature is finite the growth rate of $0.023$  from simulation agrees well with the analytical estimate obtained from 
 kinetic calculations for these parameters of $0.025$ ($ \delta^{hot}_{max}\sim \frac{2\sqrt{6}}{9\sqrt{\pi}}\frac{\left[\omega^2_{b}v^2_{0b}m_{e}/T_{b}+\omega^2_{b}(\gamma_{0b}-1/\gamma^3_{0b})\right]^{3/2}}{\omega^2_{p}(v^2_{0p}+T_{p}/m_{e})c}(T_{p}/m_{e})^{3/2}$\cite{bao}). It should be noted that this is consistent with the well known characteristic feature of the reduction in Weibel growth rate by 
 increasing  beam temperature. After the linear phase of growth, it can be observed from Fig.~\ref{fig:l_growth_Rate} that when the normalized magnetic 
 energy becomes of the order of unity the increase in  magnetic energy considerably slows down. This reflects the onset of the nonlinear regime. 
 We discuss the nonlinear regime of the instability in the next section in detail.  
\section{NONLINEAR STAGE OF INSTABILITY}
When the Weibel separated magnetic fields acquire significant magnitude, they start influencing  the dynamics of beam and plasma particles. 
This backreaction signifies the onset of nonlinear regime. The plot of the magnetic energy growth in Fig.~\ref{fig:l_growth_Rate} clearly, shows that 
at around $t \sim 50 \omega_{pe}t$ (cold beam-plasma system) the system enters the nonlinear phase. The characteristics behaviour in the nonlinear 
regime has been  described in the subsections below: 

 \subsection{Current filaments}
\begin{figure}
       \centering
               \includegraphics[width=\textwidth]{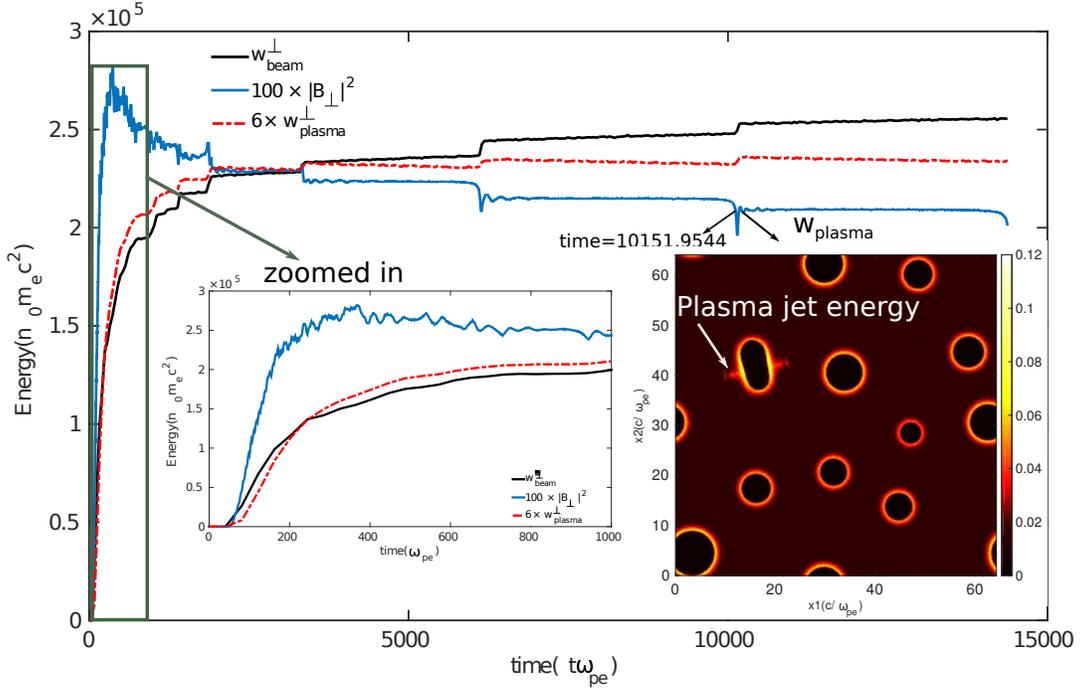}
               
                \caption {Evolution of energy ( in unit of $n_0m_ec^2$ )  with time: The solid blue curve shows the magnetic field energy $|B_{\perp}|^2=|B_x^2+B_y^2|$ and the decay in magnetic field energy at later time can be clearly seen. The solid black curve is for transverse kinetic energy $W^{\perp}_{beam}$ of beam and dotted red curve is for transverse kinetic energy $W^{\perp}_{plasma}$ of plasma. In the plots, we observe the gain in the transverse kinetic energy of plasma as well as beam at the same time when magnetic reconnection phenomena occurs and magnetic field energy decay. The color plot of kinetic energy of plasma particle has been shown in the inset of figure which shows the jet like structure.
        }
              \label{fig:nonl_growth_Rate}
             \end{figure}

In the non-linear stage of WI, the current filaments, flowing in the same direction,  merge with each other with time and organize as bigger size filaments. 
During the initial nonlinear stage magnetic field energy keeps growing, albeit at a rate which is much slower than the 
linear growth rate and then saturates (Fig.~\ref{fig:nonl_growth_Rate}). Subsequently, the magnetic field energy decreases 
as can be observed from the plot of Fig.~\ref{fig:nonl_growth_Rate}. 
 
A rough estimate of the saturated magnetic  field   can be made by the following simple consideration.  
 The spatial profile of the magnetic field in the current filament is 
  mimicked as a sinusoidal function with $k$ representation the inverse of the filament size.  The amplitude of the magnetic field is $B_0$ 
  which in the nonlinear regime significantly deflects the  trajectories  of the  electrons.
  The transverse motion of an electron in the plane of the magnetic field is then given by
\begin{equation}
 \frac{d^2r}{dt^2}=\frac{ev_{z}}{mc\gamma_{0b}}B_{0}sin(kr)
\end{equation}
The bounce frequency of a magnetically trapped electron is thus 
\begin{equation}
 \omega^2_{m}=\frac{ev_{0b}kB_{0}}{m_{e}c\gamma_{0b}}
\end{equation}
the saturation would occur when the typical  bounce  
frequency becomes equal to the maximum linear growth rate of instability. 
Therefore, the saturated magnetic field can be estimated by comparing the linear growth of filamentation instability to the bounce  frequency ($\omega_{m}=\delta_{m}$).
In case of mono-energetic distribution function, the saturated magnetic field is 
\begin{equation}
 B_{sat}\sim \left(\frac{ m_{e}v_{0b}n_{0b}}{ekcn_{0p}}\right)\omega_{p}^2
 \label{B_sat}
\end{equation}
The estimate provided by the eq.(~\ref{B_sat}) compares well with the observed saturated value of the magnetic field which is equal to 0.1.

\subsection{Alfv\'en limited filaments}
\begin{figure}
               \includegraphics[width=\textwidth]{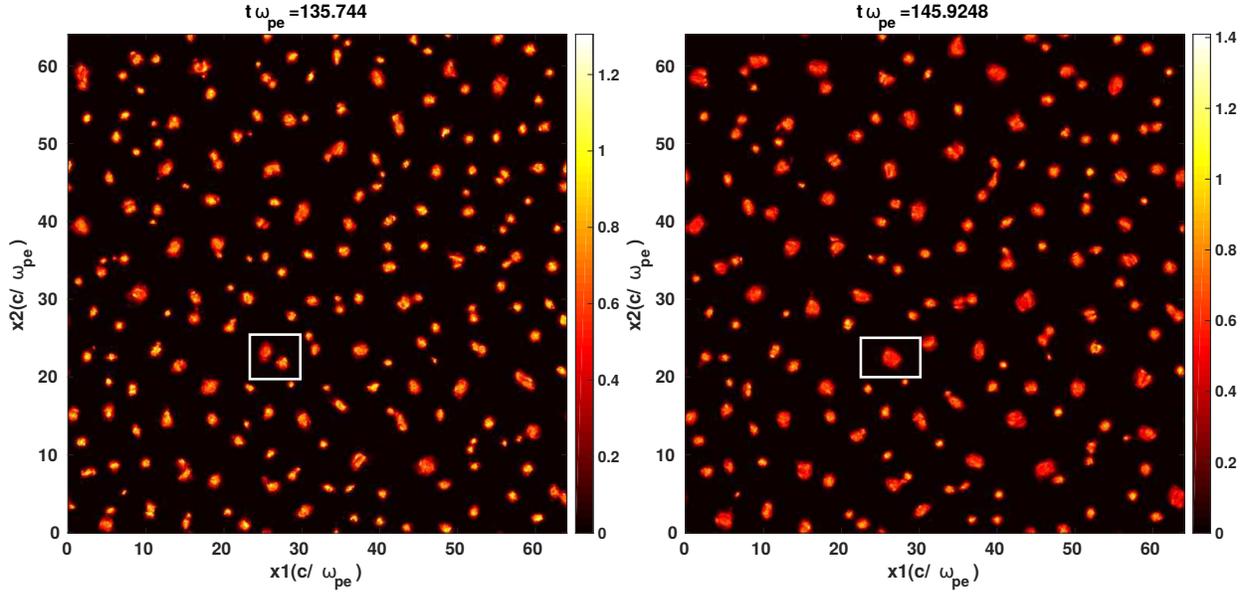}
               \caption{  The merging of beam current filaments highlighted by white box
               }
                \label{fig:evolution}
       \end{figure}

\begin{figure}
                 \includegraphics[width=\textwidth]{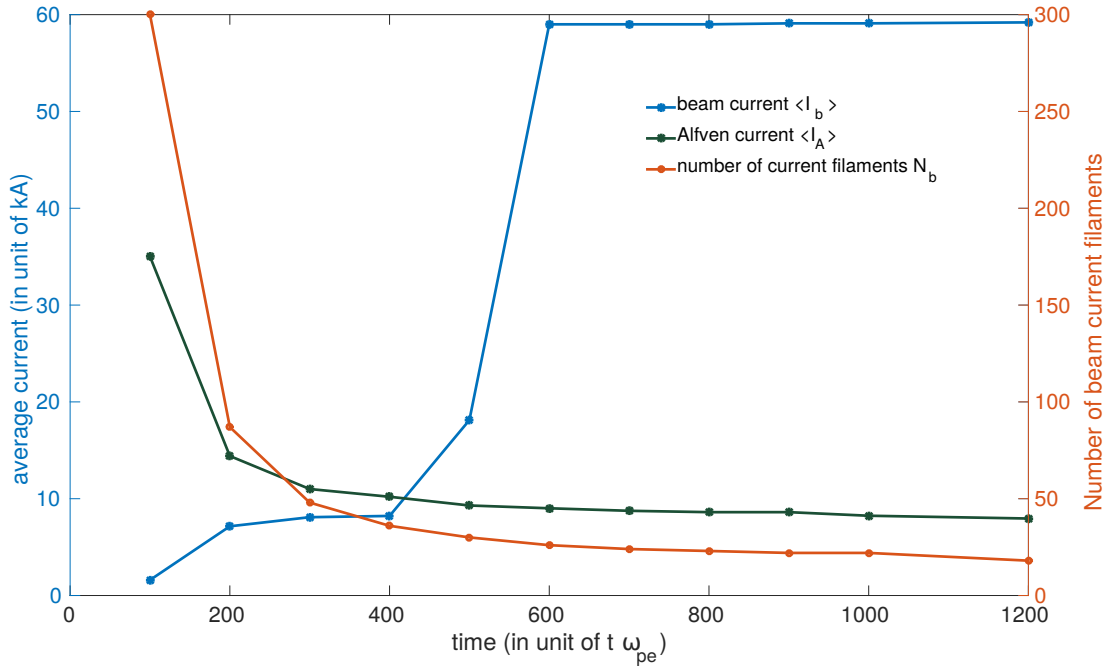} 
                 \caption{The time evolution of number of current filaments( solid red curve), average current per filament (solid blue curve) and Alfv\'en (solid green curve). }
                 \label{fig:alfven_limit}  
         \end{figure}

The process of 
merging of  like current filaments can be  seen from the plots of temporal evolution of the current densities shown in  Fig.~\ref{fig:evolution}. The current in the filament is essentially due to the beam electrons as illustrated in Fig.~\ref{fig:evolution}.
The current in the filament, however, should not exceed the  Alfv\'en limit. The value of the Alfv\'en limit for our simulations is $35 kA$. In Fig.~\ref{fig:alfven_limit}, we show the evolution of the beam current and the number of filaments in the simulation box. 

 The number of filaments keeps dropping slowly, however, the average beam current in the filament keeps increasing. Since the beam particles convert the partial kinetic energy into magnetic field energy and slow down, the Alfv\'en current limit drops with time and finally saturates. The magnetic field keep on increasing until the average beam current lower than Alfv\'en current limit and at particular time, the  average beam current cross the Alfv\'en current limit.
 This is also the time after which there is no increase in the magnetic energy of the system. After the saturation of instability,  the magnetic field energy ($|B_{\perp}|^2(n_0m_ec^2)$  zoomed $100\times$ for better view) starts decaying  
 as shown in Fig.~\ref{fig:nonl_growth_Rate}. In fact, one observes that the magnetic energy reduces after this time. This decay in the magnetic field energy can be understood on the basis of magnetic reconnection phenomena.

\subsection{Electron jet formation} 
\begin{figure}
                 \includegraphics[width=1.15\textwidth]{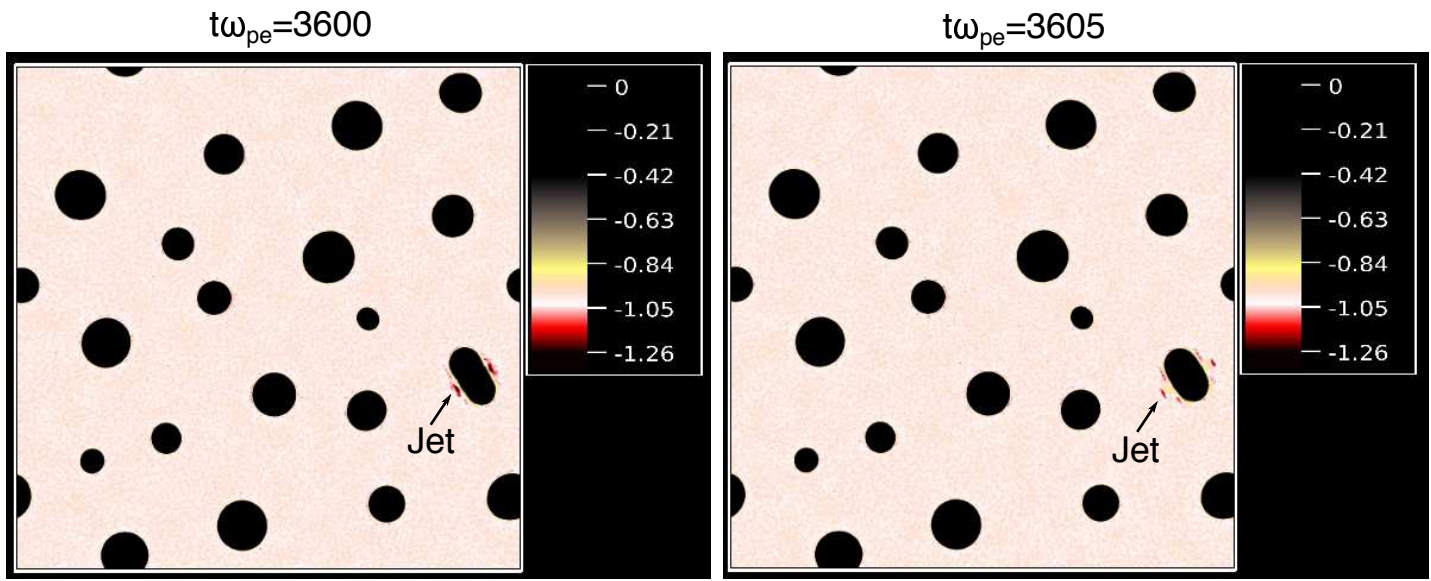} 
                 \caption{The formation of jet like structure during the magnetic reconnection process in background plasma charge density [highlighted by arrow] }
                 \label{fig:jet}  
         \end{figure}

At later times when only few filaments are left in the system, we can track each of the filaments individually. 
We choose two such filaments which are about to converge and observe their behaviour as they coalesce with each other in Fig.~\ref{fig:jet}.
The figures show the formation of electron jets in the plane as the two structures merge with each other.  

The structures basically follow the EMHD\cite{king,stenzel,das} dynamics. Thus as the filaments come near each other they carry the $\vec{B} - \nabla^2 \vec{B} $ with them. The filament scales being longer than  electron skin depth, the magnetic field is essentially carried by the electron flow in the plane. However, when the filaments hit each other 
the magnetic fields get compressed against each other and scale sizes smaller than electron skin depth are formed. The collisionless 
inertia driven reconnection takes place and generating energetic electron jets in the direction orthogonal to the direction at which the filaments approach each other. 

There is a change in 
magnetic field topology and accelerated  electrons are observed. 
This is responsible for the reduction in magnetic field energy. In 
fact, each of the sudden drops in magnetic field energy can be 
times with such reconnection events in the simulation. 
The increase in the perpendicular energy of the electrons
$W_{\perp}$ has been shown in Fig.~\ref{fig:nonl_growth_Rate}. 

It is interesting to note that  nature tries to use ingenious and rapid  techniques to relax a highly asymmetric system such as the one with  inter-penetrating 
electron current flows along $\pm \hat{z}$. The system being collisionless it would not have been possible to convert 
$W_{\parallel}$ (kinetic energy parallel to $\hat{z}$ axis to $W_{\perp}$). The magnetic field provides an intermediary role 
to aid this process and symmetrize the system. 
\section{CONCLUSIONS}
In the present article, we have studied theoretically as well as numerically (PIC simulation) the Weibel instability in beam plasma system. 
We have shown that the linear stage of  Weibel instability shows good agreement with PIC simulation results. 
In the non-linear stage, the current filaments having the flow in the same direction merge into each other.  There is a rearrangement of  the magnetic field lines through reconnection process  leading to energetic electron jets in the  plane. This has been clearly observed in the simulation. So ultimately the parallel flow energy gets converted into the perpendicular electron energy. Basically,  the 
excitation of the instability and the nonlinear evolution tries to relax the configuration towards an isotropic configuration rapidly 
in a collsionless system.

 \bibliographystyle{unsrt}

 \clearpage
%

\end{document}